\documentstyle[12pt,psfig]{article}
\newcommand{\resection}[1]{\setcounter{equation}{0}\section{#1}}

\def\abs#1{\vert #1 \vert}
\def\SRR{S_{R,R}}
\def\SRL{S_{R,L}}
\def\SLL{S_{L,L}}
\def\siso{\sigma_{{\rm iso}}}
\def\sipm{\sigma_{\pm}}
\def\tx{{\theta \over 2 \pi i}}
\def\psil{\psi^{(+)}}
\def\psir{\psi^{(-)}}
\def\lamtp{{\lambda \over 2 \pi}}
\def\lamtpi{{\lambda \over 2 \pi i}}

\def\onehalf{{1 \over 2}}
\def\state#1{\vert #1 \rangle}
\def\astate#1{\langle #1 \vert}
\def\nn{\nonumber}
\def\beq{\begin{equation}}
\def\eeq{\end{equation}}
\def\bea{\begin{eqnarray}}
\def\eea{\end{eqnarray}}
\newcommand{\ds}{\!\!\not\!\partial}
\newcommand{\As}{\not\!\! A}

\catcode`\@=11
\def\marginnote#1{}
\def\@eqnlabel{}
\def\@vacuum{}

\def\titlepage{\@restonecolfalse\if@twocolumn\@restonecoltrue\onecolumn
     \else \newpage \fi \thispagestyle{empty}\c@page\z@
        \def\thefootnote{\arabic{footnote}} }
\def\endtitlepage{\if@restonecol\twocolumn \else  \fi
        \def\thefootnote{\arabic{footnote}}
        \setcounter{footnote}{0}}  

\def\cite{\@ifnextchar[{\@tempswatrue\@citex}{\@tempswafalse\@citex[]}}

\def\@citex[#1]#2{%
\if@filesw\immediate\write\@auxout{\string\citation{#2}}\fi
\leavevmode\unskip\ \@cite{\@collapse{#2}}{#1}}

\def\@bylinecite{%
\@ifnextchar[{\@tempswatrue\@CITEX}{\@tempswafalse\@CITEX[]}%
}

\def\@CITEX[#1]#2{%
\if@filesw\immediate\write\@auxout{\string\citation{#2}}\fi
\leavevmode\unskip$^{\scriptstyle\@CITE{\@collapse{#2}}{#1}}$}

\def\@cite#1#2{[{#1\if@tempswa , #2\fi}]} %
\def\@CITE#1#2{{#1\if@tempswa , #2\fi}} %

\def\@collapse#1{%
{%
\let\@temp\relax
\@tempcntb\@MM
\def\@citea{}%
\@for \@citeb:=#1\do{%
\@ifundefined{b@\@citeb}%
{\@temp\@citea{\bf ?}%
\@tempcntb\@MM\let\@temp\relax
\@warning{Citation `\@citeb ' on page \thepage\space undefined}%
}%
{\@tempcnta\@tempcntb \advance\@tempcnta\@ne
\edef\MyTemp{\csname b@\@citeb\endcsname}%
\def\@tempa{\@temptokena=\bgroup}%
\afterassignment\@tempa %
\@tempcntb=0\MyTemp\relax}%
\ifnum\@tempcntb=0\relax%
\@tempcntb=\@MM
\@citea\MyTemp
\let\@temp = \relax
\else %
\edef\@tempd{\number\@tempcntb}%
\ifnum\@tempcnta=\@tempcntb %
\ifx\@temp\relax %
\edef\@temp{\@citea\@tempd}%
\else
\edef\@temp{\hbox{--}\@tempd}%
\fi
\else %
\@temp\@citea\@tempd
\let\@temp\relax
\fi
\fi
}%
\def\@citea{,}%
}%
\@temp %
}%
}%

\advance\textheight by 5mm
\advance\textwidth by 4mm
\begin{document}
\setcounter{page}{0}
\begin{titlepage}
\begin{center}
May~1997 \hfill hep-th/9705180 \\
$~$ \hfill IC/97/37   \\
$~$ \hfill SISSA 54/97/EP \\
[.47in]
{\large \bf A Non-Perturbative Approach to \\ 
[0.03in]
           the Random-Bond Ising Model}\\
[.5in] {\bf
   D.~C.~Cabra$^{a,}$\footnote{Investigador CONICET, Argentina. On 
                      leave from Universidad Nacional de la Plata, Argentina.}, 
   A.~Honecker$^{b,}$\footnote{Work done under support of the EC TMR Programme
             {\em Integrability, non-per\-turba\-tive effects and symmetry in
             Quantum Field Theories}, grant FMRX-CT96-0012}, 
   G.~Mussardo$^{a,b,c,2}$ and P. Pujol$^b$}
\vspace{7mm} \\ 
        $^a${\it International Centre for Theoretical Physics,  
        34100 Trieste, Italy}
\vspace{4mm} \\
        $^b${\it International School for Advanced Studies,
        34014 Trieste, Italy}
\vspace{4mm} \\  
        $^c${\it Institute of Theoretical Physics, University of California\\
        Santa Barbara 96106, U.S.A.} 
\end{center}

\vskip .5in
\centerline{\bf ABSTRACT}
\begin{quotation}
We study the $N \rightarrow 0$ limit of the $O(N)$ Gross-Neveu model in the 
framework of the massless form-factor approach. This model is related to 
the continuum limit of the Ising model with random bonds via the replica
method. We discuss how this method may be useful in calculating correlation
functions of physical operators. The identification of non-perturbative
fixed points of the $O(N)$ Gross-Neveu model is pursued by its mapping to a
WZW model. 
\vskip 0.5cm 

\end{quotation}
\end{titlepage}
\newpage

\resection{Introduction}

The aim of this paper is to discuss the critical regime of the
two-di\-men\-sional random bond Ising model by applying
non-perturbative methods to the massless 
phase of the $O(N)$ Gross-Neveu (GN) model in the limit $N \rightarrow 0$. It 
is well known (and briefly sketched below) that in the context of the 
replica method the analytic continuation $N \rightarrow 0$ of the GN model 
describes the quenched averages of the Ising model in the presence of 
gaussian distributed random couplings \cite{dots1,shlaev,shankar,ludwig2,DPP}. 
Perturbative calculations based on the GN model have then been extensively and 
successfully used for studying the behavior of correlation functions of the 
random model in the infrared regime where the GN model for $N < 2$ is
asymptotically free \cite{shlaev,ludwig2,DPP}. Being an asymptotically free
infrared theory, the perturbative series is plagued though by the presence of 
Landau poles which do not permit to study the behavior of the correlators 
on their short distance scales and therefore to follow in general the change 
of the theory in passing from the short to the large distance scales. In this 
paper we show how in some cases we may get around this problem by using 
the non-perturbative massless Form-Factor (FF) approach proposed in \cite{DMS}.  
For the two-point function of the energy operator of the random model we show 
for instance that the absence of Landau poles in the $S$-matrix formulation 
allows us to have much more information on the correlator at intermediate 
(and small) scales than what can be obtained within a perturbative 
renormalization group approach. A more general theoretical problem 
consists in identifying the ultraviolet fixed point of the GN model in 
its entire massless range $N < 2$. At the end of this paper we 
present some considerations on a mapping of the GN model to an interacting 
Wess-Zumino-Witten (WZW) model which may be useful for further 
investigation of this problem. 

\resection{$S$-matrix in replica space} 

Let us begin our discussion with the continuum limit of the random-bond 
Ising model expressed in terms of a Majorana fermion (see for example 
\cite{dots1} for details) with the partition function given by 
\beq
{\cal Z}[m(x)] = \int D[\psi] ~\exp\left[ -\int d^2x ~ \overline{\psi} ~
(~\ds -
m(x)) ~\psi \right] \,\,\, .
\label{zz}
\eeq
In the above formula, $m(x)$ is a (position dependent) random mass term 
associated to the lattice random bond interactions. We assume that $m(x)$ 
has a gaussian distributed probability 
$P[m(x)] \propto \exp \left(-m^2(x) / 4 g \right)$. Let us now make 
use of the standard replica method to compute 
the average of the free energy:
\beq
\overline{\ln{\cal Z}} =
   \lim_{N \rightarrow 0} \overline{{\cal Z}^N -1 \over N} \,\,\, .
\label{zzz}
\eeq
The quenched averages of correlation functions can then be obtained by 
adding a source term to the partition function (\ref{zz}) and differentiating 
(\ref{zzz}) with respect to it. The above procedure leads to an effective
action described by the Gross-Neveu model with $N \rightarrow 0$ colors:
\beq
{\cal S} = \int d^2x \left[ \overline{\psi}_a \ds  \psi_a 
+ g \left(\overline{\psi}_a  \psi_a \right)^2 \right] \, ; \qquad
a=1, \cdots , N \, .
\label{GNlag}
\eeq
Irrespective of the number of colors $N$, the GN model presents
at the classical level an infinite number of local and non-local
charges which are conserved just in virtue of the equations of
motion \cite{Neveu,Luscher}. 
For $N > 2$, the integrability at the classical level is known to survive at 
the quantum level: the associated quantum field theory is asymptotically free 
in the ultraviolet region, massive \cite{GN} and presents in general quite 
a rich spectrum of bound states. The integrability of the quantum model 
implies the elasticity and factorization of its $S$-matrix amplitudes which 
can be then determined by employing the unitarity and crossing symmetry 
equations, together with the residue equations coming from the bootstrap 
principle \cite{Karowski,ZamZam}. 

For $N<2$ the model is instead asymptotically free in the infrared region and 
presents at the quantum level a massless phase, as it can be argued 
(at least perturbatively) from the changing of the sign of the $\beta$-function 
near $g=0$ \cite{Wetzel}. Its spectrum consists in this case of only $N$ 
massless Majorana fermions which can be right- or left-movers with a dispersion 
relation parameterized in terms of a crossover scale $M$ and the rapidity 
variable as $E=p=(M/2) ~\exp(\theta)$ and $E=-p=(M/2) ~\exp(-\theta)$ 
respectively \cite{MuSi}. Assuming that the integrability\footnote{This 
statement can be actually proven along the arguments presented in \cite{Witten} 
if one is inclined to be rather cavalier about the meaning of the color 
indices in this range of $N$.} of the model also holds for $N<2$, we propose 
the following exact $S$-matrix involving its fermionic massless 
excitations (for the general formulation of massless scattering theories, see 
\cite{massless}) \footnote{This essentially follows the original proposal
made in \cite{MuSi} for the massless $S$-matrix of the GN model, with the
difference that there the $S$-matrix in the $L-L$ and $R-R$ channel was
incorrectly assumed to be $-1$ on the basis of Feynman rules derived from
the Lagrangian (\ref{GNlag}) which however involves for $N < 2$
irrelevant non-renormalizable interactions.}  
\beq  
\SRR = \SLL = \SRL = S
\label{equalS} 
\eeq 
where 
\beq  
S_{a,b}^{c,d}(\theta) =
  \delta_{a,b} \delta^{c,d} \sigma_1(\theta)
+ \delta_{a}^{c} \delta_{b}^{d} \sigma_2(\theta)
+ \delta_{a}^{d} \delta_{b}^{c} \sigma_3(\theta)
\label{S-chans} 
\eeq 
with
\bea 
\sigma_1(\theta) &=& - {i \lambda \over i \pi - \theta} \sigma_2(\theta)
\,\, , \qquad\qquad
\sigma_3(\theta) = - {i \lambda \over \theta} \sigma_2(\theta) \,\, , \nn \\
\sigma_2(\theta) &=& - {
\Gamma(1 - \tx) \Gamma(\onehalf + \tx) \Gamma(\onehalf - \lamtpi - \tx)
  \Gamma(- \lamtpi + \tx) \over
\Gamma(\tx) \Gamma(\onehalf - \lamtpi + \tx) \Gamma(\onehalf - \tx)
  \Gamma(1- \lamtpi - \tx)} 
\label{defSi} 
\eea 
and
\beq 
\lambda = {2 \pi \over N-2} \, .
\label{valLam} 
\eeq 
The above expressions formally coincide with those discussed
in \cite{ZamZam} for the massive phase of the model, the main distinctions 
being the different role played by the rapidity parameter 
as well as the different range of the parameter $N$ in the two cases. 
The ansatz (\ref{defSi}) is supported by the validity of the 
Yang-Baxter equations for all possible combinations of right- and left-moving 
particles as well as by the $O(N)$-invariance of the interaction. Notice that 
for $N < 2$ the above amplitudes do not have poles in the physical sheet (a 
feature which is consistent with the massless phase of the model). For 
$N \rightarrow 2$ they are continuous functions and at $N=2$ reduce to the
$S$-matrix amplitudes of the Sine-Gordon model at the Coleman transition point
$\beta^2 \rightarrow 8 \pi$
\bea
\sigma_2(\theta) &=& 0 \,\, , \nn \\
\sigma_1(\theta) &=& {-2i\pi \over i\pi - \theta}
    ~{\Gamma(1 - \tx) \Gamma(\onehalf + \tx)
       \over \Gamma(\tx) \Gamma(\onehalf - \tx)}
\,\, , \\
\sigma_3(\theta) &=& {-2i\pi \over \theta}
    ~{\Gamma(1 - \tx) \Gamma(\onehalf + \tx)
       \over \Gamma(\tx) \Gamma(\onehalf - \tx)}
\,\, .  \nn
\eea

\resection{Minimal Form-Factors}

Let us consider now the calculation of correlation functions by means of 
the spectral representation based on the FFs (see for instance 
\cite{DMS,KaWei,Smirnovbook} for the relevant formulas and notation). 
For the two-particle FF the problem can be initially reduced to consider the 
three different channels whose $S$-matrices are given by:
\beq 
\siso = N \sigma_1 + \sigma_2 + \sigma_3 \, , \qquad
\sipm = \sigma_2 \pm \sigma_3 \,\,\, .
\label{defSiso} 
\eeq 
For each of these channels, the minimal FF can be evaluated. As an example we 
construct the one for the isoscalar channel. Watson's equations for a massless 
flow are \cite{DMS} (see also \cite{MeSm}):
\beq 
F_{\alpha_1, \alpha_2}(\theta)
  = S_{\alpha_1, \alpha_2}(\theta) F_{\alpha_2, \alpha_1}(-\theta)
\, , \qquad
F_{\alpha_1, \alpha_2}(\theta + 2 \pi i) = F_{\alpha_2, \alpha_1}(-\theta)
\label{WatsonEq} 
\eeq 
where $\alpha_i = R, L$ and $S$ stands for a scalar $S$-matrix which in our 
case is $\siso$. In order to solve these equations it may be useful to 
employ an integral representation for $\siso$:
\beq 
\siso(\theta) = -
\exp\left\{-2 \int\limits_{0}^{\infty} {{\rm d} x \over x} \,
{{\rm e}^{({\lambda \over \pi} + 1) x} - {\rm e}^{-x} \over
{\rm e}^x + 1} \sinh\left({\theta x \over i \pi }\right) 
\right\} \, .
\label{IntRepSiso} 
\eeq 
(Note that for $N=1$ this simply reduces to $-1$ -- the $S$-matrix
of the Ising model). Let us write $ F_{\alpha_1, \alpha_2}(\theta)$ as  
\beq 
F_{\alpha_1, \alpha_2}(\theta) 
= f_{\alpha_1, \alpha_2}(\theta)
g_{\alpha_1, \alpha_2}(\theta)
\label{factorFF} 
\eeq 
where $g_{\alpha_1, \alpha_2}(\theta)$ satisfies 
\bea
g_{\alpha_1, \alpha_2}(\theta) &=& - g_{\alpha_2, \alpha_1}(-\theta)
\, , \qquad
g_{\alpha_1, \alpha_2}(\theta + 2 \pi i) = g_{\alpha_2, \alpha_1}(-\theta)
\, , \label{g-Eq} 
\eea
whereas $f_{\alpha_1, \alpha_2}(\theta)$ fulfills the equations 
\bea
&& f_{\alpha_1, \alpha_2}(\theta) = 
\exp\left\{-2 \int\limits_{0}^{\infty} {{\rm d} x \over x} \,
{{\rm e}^{({\lambda \over \pi} + 1) x} - {\rm e}^{-x} \over
{\rm e}^x + 1} \sinh\left({\theta x \over i \pi }\right)
\right\} f_{\alpha_2, \alpha_1}(-\theta)
\,\, , \nn \\
&& f_{\alpha_1, \alpha_2}(\theta + 2 \pi i) =\, f_{\alpha_2, 
\alpha_1}(-\theta)
\,\,\, . \label{f-Eq}
\eea 
The solution of (\ref{g-Eq}) for the $R-R$ and $L-L$ channels is unique
(up to a normalization constant) and is given by 
\beq 
g_{R,R}(\theta) = g_{L,L}(\theta) = \sinh\left({\theta \over 
2}\right) \, .
\label{RRLLg} 
\eeq 
For the $R-L$ channel any linear combination of ${\rm e}^{\theta \over 2}$
and ${\rm e}^{-{\theta \over 2}}$ is a solution. The exact computation
of the two-particle FF for the energy operator for $N=1$ (which corresponds
to the pure Ising model) and its expected infrared behavior at $N=0$ fixes
the solution in a unique way
\beq 
g_{R,L}(\theta) = {\rm e}^{\theta \over 2} \, .
\label{RLg} 
\eeq 
As for (\ref{f-Eq}) we take the same solution in the $R-R$, $L-L$ and
$R-L$ channels (this assumes that $f_{R,L}(\theta) = f_{L,R}(\theta)$):
\beq 
f_{R,R} = f_{L,L} = f_{R,L} = f \,\, ,
\label{setFeq} 
\eeq 
with the function $f$ given by 
\beq 
f(\theta) =
\exp\left\{-2 \int\limits_{0}^{\infty} {{\rm d} x \over x} \,
{{\rm e}^{({\lambda \over \pi} + 1) x} - {\rm e}^{-x} \over
{\rm e}^x + 1} \, {\sin^2\left({x (i \pi - \theta) \over 2 \pi }\right)
\over \sinh{x}} \right\} \,\,\,. 
\label{fIntRep} 
\eeq 
This can be also expressed as 
\bea
&& f(\theta) = \prod_{k=0}^{\infty} \left(
{\Gamma(1-\lamtp+k)\Gamma({3 \over 2} +k) \over
\Gamma(2+k) \Gamma(\onehalf-\lamtp+k)}\right)^2 
\label{fGamRep} \\
&& \ \times {\Gamma({5 \over 2} - \tx + k) \Gamma({3 \over 2} + \tx + k)
\Gamma(1 - \lamtp - \tx + k) \Gamma(- \lamtp  + \tx + k)
\over
\Gamma(2 - \tx + k) \Gamma(1 + \tx + k)
\Gamma({3 \over 2} - \lamtp - \tx + k) \Gamma(\onehalf - \lamtp  + \tx + k)}
\, .\nn
\eea
Using its infinite-product representation one can show that $f(\theta)$ 
satisfies the following functional relation: 
\beq 
f(\theta) f(\theta+i \pi) = {\cal C}_{\lambda}
{\Gamma(- \lamtp  + \tx) \Gamma(\onehalf - \lamtp  - \tx) \over
\Gamma(1 + \tx) \Gamma({3 \over 2} - \tx)} \, ,
\label{functRel} 
\eeq 
which may be useful in the computation of higher-particle 
FFs.
The constant ${\cal C}_{\lambda}$ is given by
\beq 
{\cal C}_{\lambda} = {\Gamma({3 \over 2}) \over
\Gamma(- \lamtp) \Gamma(\onehalf - \lamtp)} 
\prod_{k=0}^{\infty} \left(
{\Gamma(1-\lamtp+k)\Gamma({3 \over 2} +k) \over
\Gamma(2+k) \Gamma(\onehalf-\lamtp+k)}\right)^4 \!
{(1 + k) ({3 \over 2} +k) \over (-\lamtp +k) (\onehalf-\lamtp+k)}
\, ;
\label{valFconst} 
\eeq
its numerical value at $N=0$ is ${\cal C}_{- \pi} = 0.5854\ldots$.

\resection{Two--point function of the energy operator}

Let us specialise our analysis to evaluate the (average) correlation function
of the energy operator $\overline{\langle\epsilon(x) \epsilon(0)\rangle}$. In
terms of the replica, this is equivalent to evaluate the correlation function 
\beq
\frac{1}{N} \sum_{k=1}^N \left.
\langle \epsilon_k(x) \epsilon_k(0)\rangle \right\vert_{N=0}
\label{replicacorr}
\eeq
in the GN model. At the infrared fixed point, the energy operator is given in
terms of the fermions by $\epsilon_k(x) \sim -i \psir_k(x) \psil_k(x)$
where $\psir_k(x)$ and $\psil_k(x)$ are the chiral components of the original
Majorana fermions. Duality and spin reversal symmetry translate into
invariance of the Lagrangian (\ref{GNlag}) under the two transformations
$\psil_k \mapsto - \psil_k$ and $\psir_k \mapsto \psir_k$
or $\psil_k \mapsto \psil_k$ and $\psir_k \mapsto -\psir_k$.
Under these two transformations, the energy operator changes its sign. Since
these discrete symmetries are preserved by the perturbation, one concludes
that the only non-vanishing FFs of the energy-operator are those with an 
odd number of both left- and right-moving particles. In particular, the first 
non-trivial FF is
\beq 
F_{i,j}^{R,L}(\theta_1,\theta_2) = \astate{0} 
\left(\sum_{k=1}^N \epsilon_k(0)\right)
\state{R_i(\theta_1) L_j(\theta_2)} \, .
\label{defFFenOp} 
\eeq 
Here $\state{R_i(\theta_1) L_j(\theta_2)}$ corresponds to an asymptotic
state of one right-moving and one left-moving fermionic
particle. $O(N)$ and Lorentz invariance fixes this FF (up to an overall 
normalization constant) to be given by
$F_{i,j}^{R,L}(\theta_1,\theta_2)= \delta_{i,j} F_{R,L}(\theta_1 - \theta_2)$
with $F_{R,L}$ as in eqs.\ (\ref{factorFF}), (\ref{RLg}), (\ref{fIntRep}).
In the previous expression there could be in principle a supplementary factor
(a $2 \pi i$ periodic symmetric function without poles, {\it i.e.}\ a
polynomial in ${\rm e}^{\theta_i}$). This factor is actually absent in the 
pure Ising model ($N=1$) and its absence will be further justified
later on by the correct asymptotic behavior of the correlation function 
(\ref{replicacorr}). The contribution of (\ref{defFFenOp}) to the two-point
correlation function of the energy operator is given by 
\bea 
\langle \epsilon(x) \epsilon(0) \rangle^{(2)} &\sim&
\int\limits_{-\infty}^{\infty} {{\rm d} \theta_1 \over 2 \pi}
\int\limits_{-\infty}^{\infty} {{\rm d} \theta_2 \over 2 \pi}
\ \abs{F_{R,L}(\theta_1 - \theta_2)}^2 \
\exp\left(-{M r \over 2}
\left[{\rm e}^{\theta_1} + {\rm e}^{-\theta_2}\right]\right) \nn \\
&\equiv & C_{\epsilon}^{(2)}(Mr)
\label{corTwoPart} 
\eea 
where the superscript indicates that this is the two-particle contribution and 
$x=(ir,0)$ \footnote{ Actually, there is another two-particle contribution to
this correlation function, namely the one coming from $F_{L,R}$. However,
it gives the same contribution as $F_{R,L}$.}. With the change of 
variables $\gamma = \theta_1 + \theta_2$, $\theta = \theta_1 - \theta_2$
we obtain
\bea 
C_{\epsilon}^{(2)}(Mr)
&=&
{1 \over 2}
\int\limits_{-\infty}^{\infty} {{\rm d} \theta \over 2 \pi}
\int\limits_{-\infty}^{\infty} {{\rm d} \gamma \over 2 \pi}
\ \abs{F_{R,L}(\theta)}^2 \
\exp\left(-M r \; {\rm e}^{{\theta \over 2}} \cosh{\gamma\over 2}\right)
\nn \\
&=&
\int\limits_{-\infty}^{\infty} {{\rm d} \theta \over 2 \pi^2}
 K_0(M r \; {\rm e}^{{\theta \over 2}})
\ \abs{F_{R,L}(\theta)}^2 \, . 
\label{CorK0exp} 
\eea 
Using (\ref{factorFF}), (\ref{RLg})--(\ref{fIntRep}) we can express
$\abs{F_{R,L}(\theta)}^2$ as (from now on we will specialize our formulas 
to the limit $N \to 0$) 
\beq 
\abs{F_{R,L}(\theta)}^2 =
{{\rm e}^{\theta} \over \sqrt{1 + {\theta^2 \over \pi^2}}}
\, \Phi(\theta)
\label{factFRL} 
\eeq 
with
\beq 
\Phi(\theta) = \exp\left\{2
\int\limits_{0}^{\infty} {{\rm d} x \over x} {\rm e}^{-x} \,
\tanh^2\left({x \over 2}\right) \cos^2\left({x \theta \over 2 \pi}\right)
\right\} \, .
\label{repPhiInt} 
\eeq 
It is straightforward to evaluate this last integral numerically,
it is bounded from below by 0 and from above by
\beq 
2 \int\limits_{0}^{\infty} {{\rm d} x \over x} {\rm e}^{-x} \,
\tanh^2\left({x \over 2}\right)
 = 0.315384\ldots \, ,
\eeq 
{\it i.e.}\
\beq 
1 \le \Phi(\theta) \le 1.370786
\label{boundsPhi} 
\eeq 
for any value of $\theta$. In particular, the factor $\Phi(\theta)$
will not influence the $\theta \to \pm \infty$ asymptotics of the FF.
Thus, we can replace $\Phi(\theta)$ by a constant in order to discuss
the infrared behavior of the contribution (\ref{CorK0exp}) to the
correlation function. Putting
\beq 
\Phi(\theta) = 1 \, ,
\label{firstApprox} \eeq 
we find
\beq 
C_{\epsilon}^{(2)}(Mr) = \int\limits_0^{\infty} {{\rm d} p \over 
\pi^2} \;
{p \over \sqrt{1 + 4 {\ln^2{p} \over \pi^2}}} \; K_0(M r \; p) \, ,
\qquad\qquad M r \gg 1 
\label{approxCeps} 
\eeq 
where we have made the change of variables $p = {\rm e}^{{\theta \over 
2}}$. Upon expansion of the integrand, the leading infrared behavior is 
found to be given by 
\beq 
C_{\epsilon}^{(2)}(Mr) = {1 \over 2 \pi (M r)^2 \ln{M r}}
\left(1+ O\left({1 \over \ln{M r}}\right)\right) \,\,\, .
\label{IRasymCeps} 
\eeq 
\begin{figure}[t]
\psfig{figure=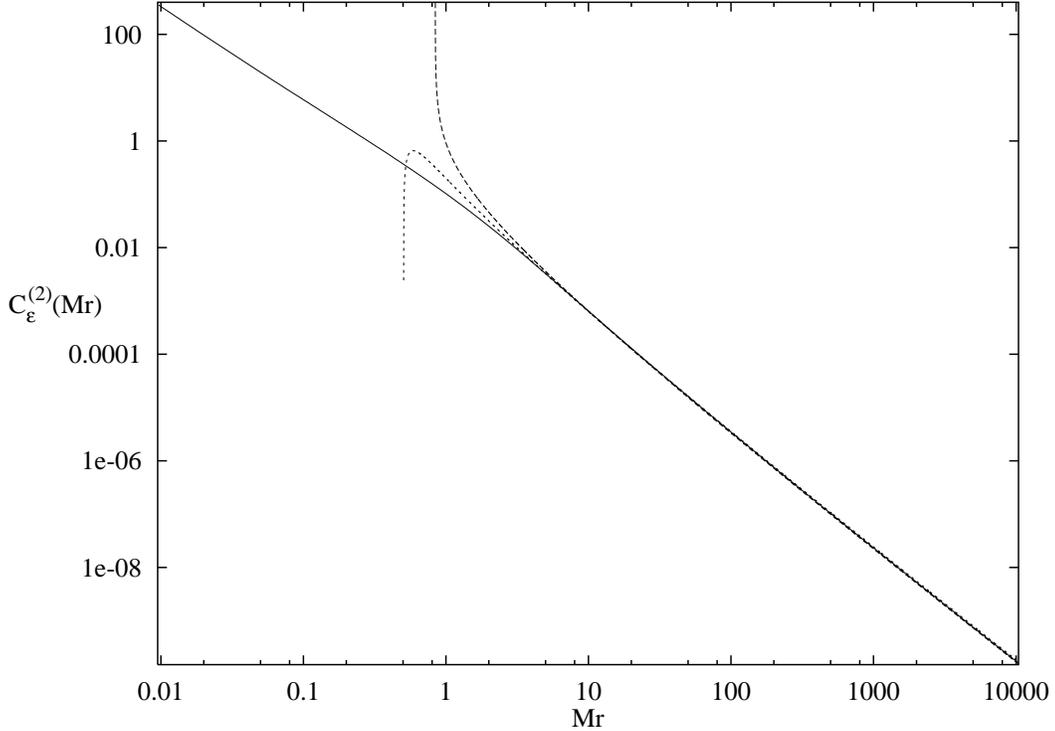,width=\columnwidth,angle=270}
\caption{
Two-particle approximation to the correlation
function of the energy operator (full line),
the one-loop perturbative result (long dashes)
and the two-loop result (short dashes).
\label{figure1}
}
\end{figure}
This fits nicely with renormalization group results, as the following
discussion shows. At the one-loop order, we have (see e.g.\ 
eq.\ (4.10) of ref.\,\cite{DPP})
\beq 
\langle \epsilon(x) \epsilon(0) \rangle =
r^{-2} \; {4 g_0 \over 1 + 8 \pi g_0 \ln{r}} + {\cal O}(g_0) 
\label{oneLoop} 
\eeq 
and including the two-loop contribution one arrives at
\bea 
\langle \epsilon(x) \epsilon(0) \rangle &=&  r^{-2} \;
{4 g_0 \over (1 + 8 \pi g_0 \ln{r}) (1 - 2 \pi g_0)}
\label{twoLoop} \times \\
&& \quad \left\{ 
1 + {4 \pi g_0 \over 1 + 8 \pi g_0 \ln{r}} \left[
\ln\left(1 + 8 \pi g_0 \ln{r}\right) - 4 \pi g_0 \ln{r} \right]
\right\}  
+ {\cal O}(g_0^2) \nn
\eea 
where $g_0$ is a bare coupling constant. Here we have chosen the
normalization such that both (\ref{oneLoop}) and (\ref{twoLoop})
agree with (\ref{IRasymCeps}) at $M=1$ for $r \to \infty$. Setting
$M=1$ we obtain a good agreement with (\ref{CorK0exp}) in the region
$M r \ge 10$ for (\ref{oneLoop}) with $g_0 = 0.23$ and for (\ref{twoLoop})
with $g_0 = 0.038$. One can see that the Landau pole in the 
two-loop formula is shifted towards smaller scales and the behavior of 
(\ref{twoLoop}) is more regular around $M r = 1$ in Fig.\ 1. Higher loop
corrections in the pertubative framework are expected to move this pole
to smaller and smaller scales. The main point is the absence of this kind
of singularity in our result based on the FF approach.
 

\resection{Comments on higher-particle FFs}

Apart from logarithmic corrections, the energy operator has canonical scaling  
dimensions in the infrared and, as explicitly checked for similar operators  
in other massless scattering theories, a rapidly convergent spectral 
series for its correlation functions is expected in this case \cite{DMS,LSS}. 
This expectation seems indeed confirmed by the above calculation. Under these 
favorable circumstances, one is essentially released from computing the more 
complicated and cumbersome expressions of the higher-particle FFs for most 
reasonable purposes. It should be pointed out though, that these favorable 
cases do not always occur. Consider, for instance the (averaged) two-point 
correlation function of the spin operator. In this case two kinds of technical 
problems arise: the first difficulty is that the integrals entering the spectral 
series ought to be regularised in order to cure the infrared divergencies. This is 
however a well-known problem of the spectral representation based on massless FFs 
\cite{DMS}, a problem  that now can be easily handled following the suggestion of 
ref.\,\cite{lessal}. The second difficulty is the most serious one,
namely the apparent necessity of 
employing in this case all higher-particle FFs of the spin operator in order to 
recover both its infrared and ultraviolet behavior. The problem is particularly 
difficult here since there is a huge difference in computing higher-particle FFs 
in diagonal rather than in non-diagonal scattering theories (in the diagonal case one 
is often able to obtain a closed expression for all the higher-particle 
FFs, see for instance \cite{ShG}). Although progress has been recently
achieved on the FF problem in some non-diagonal scattering theories
\cite{MeSm,Kar}, the general determination of the higher-particle FFs for
non-diagonal $S$-matrix models still remains an open problem of a (quite)
mathematical difficulty. It would be clearly interesting to try to develop
the massless FF approach further in order to deal successfully with this
class of operators. 


\resection{Non-trivial fixed points}
 
In the last part of this paper, we present some considerations which may be 
useful to better understand the structure of the fixed points in the $O(N)$ GN 
model and in particular to determine its ultraviolet fixed point in the 
range $N < 2$. The theory behaves once again differently
for $N > 2$ and for $N < 2$. More is known about the structure of the 
fixed points of the former case rather than of the latter. Let us in fact
recall that for $N >2$ the model is 
massive but ultraviolet asymptotically free. This means that at short 
distances it reduces to $N$ free Majorana fields (with central charge 
$c_{{\rm uv}} = N/2$) whereas at large distance scales no massless degrees of 
freedom are left and its central charge is therefore $c_{{\rm ir}} = 0$.
Conversely, for $N < 2$ the model is asymptotically free in its infrared
scales and there we have correspondingly $c_{{\rm ir}} = N/2$. However, since
in this case the model is massless along all its flow from large to short
distance scales, the central charge in the ultraviolet limit is one
of the dynamical data which remain to be determined. Some insight can be
gained by rewriting the partition function of the model defined by the action
(\ref{GNlag}) in a way more appropriate for studying the existence of
some other fixed point in addition to the (trivial) one at $g=0$ (a similar
procedure has been used in \cite{MS} to study non-trivial fixed points in
the chiral $SU(N)$ GN model). 
To this end, we first make use of the identity 
\beq
\left( \bar \psi_a \psi_a \right)^2=-2\left( \bar \psi_a 
\gamma_{\mu}T^A_{ab}\psi_b \right)^2,
\label{a}
\eeq
where $T^A$ are the $O(N)$ generators normalized as ${\rm tr} 
T^AT^B=\delta^{AB}$ (this identity is a direct consequence of the fact that 
we are working with Majorana fermions). Then the quartic term in the 
interaction can be traded for a quadratic term via the introduction of an 
auxiliary field through the identity 
\bea
&& \exp \left(2g  \int d^2 x \left( \bar \psi_a 
\gamma_{\mu}T^A_{ab}\psi_b \right)^2
\right) = \nn \\
&& \qquad \int DA_{\mu}^A \exp \left(\frac{1}{8g}\int d^2 x (A_{\mu}^A)^2
+ i\int d^2x  (\bar \psi 
\gamma_{\mu}T^A\psi) A_{\mu}^A \right),
\label{b}
\eea
and hence the partition function can be written as
\beq
{\cal Z}=\int D\bar \psi_a D\psi_a  DA_{\mu}^A 
\exp \left[ -\int d^2 x 
\left( \bar \psi_a(\ds \delta_{ab}+i\As_{ab})\psi_b+
 \frac{1}{8g} A_{\mu}^A A_{\mu}^A\right) \right] \, .
\label{c}
\eeq
The key observation now is that the fields in eq.\ (\ref{c}) can be decoupled 
through the transformations
\beq
A = -\partial h h^T \, , \qquad
\bar A = -\bar \partial g g^T \, , \qquad
\psi =\pmatrix{g & 0\cr
             0 & h\cr} \chi \, , 
\label{d}
\eeq
where $h$ and $g$ are $O(N)$ matrix-valued fields, and ${}^{T}$ stands 
for transpose. Taking into account the Jacobians of the above transformations 
\cite{FiPol} (for more details compare also Section 9
of \cite{BRS}), we can rewrite the whole partition function as a product of 
decoupled sectors
\beq
{\cal{Z}}={\cal{Z}}_{{\rm ff}}{\cal{Z}}_{{\rm gh}}{\cal{Z}}_{{\rm int}} \, ,
\label{e}
\eeq
where ${\cal{Z}}_{{\rm ff}}$ is the partition function for the $N$ 
free Majorana fermions, ${\cal{Z}}_{{\rm gh}}$ the ghost partition function 
arising in the computation of the Jacobians and
\bea
{\cal{Z}}_{{\rm int}}&=&\int Dh Dg \exp\left\{(1+2C_v)\left(
\Gamma [h]+
\Gamma [g^T]\right)\right.
\nonumber \\
&-& \left(\frac{(1+2C_v)\alpha}{2 \pi}+ \frac{1}{8g}\right) \int d^2 x\;
{\rm tr}\left(\left.\partial h h^T \bar \partial g g^T\right)
\right\},
\label{f}
\eea
where $\alpha$ is a parameter that keeps track of regularization ambiguities,
$C_v$ is the dual Coxeter number of $O(N)$ and $\Gamma [u]$ is the WZW 
action \cite{Wi}. Hence, modulo decoupled conformally invariant sectors, we 
have an effective theory of interacting WZW fields.

One comment is in order regarding the regularization ambiguities 
arising in the evaluation of the Jacobians. The quadratic term in 
$A_{\mu}^A$ in (\ref{c}) breaks both gauge invariance and local chiral 
invariance, hence there is no a priori reason to choose a regularization 
preserving any of these two symmetry transformations.
We will fix the parameter $\alpha$ later on. Using the Polyakov-Wiegmann
identity \cite{PW2} it is straightforward to see the existence of a 
fixed point given by the value of the coupling constant 
$g_1 =\pi / (4(1+2C_v)(1-\alpha))$. The effective partition function at
this value may be written as
\beq
\left. {\cal{Z}}_{{\rm int}}\right\vert_{g_1}  =
\int D\tilde g \exp\left((1+2 C_v )  \Gamma[\tilde g] \right) ,
\label{i}
\eeq
where the identification $\tilde g=g^Th$ has been made and the integral 
over $h$ has been factored out. Taking into account the free fermions and the 
ghosts, the partition function (\ref{e}) corresponds to a conformal field
theory whose Virasoro central charge is given by 
\beq
c_1 = {N \over 2} - N(N-1) +  \left[{(1+2C_v)N(N-1) \over
2(1+C_v)}\right] 
= \frac{N}{2} \left(1- \frac{N-1}{1 + C_v}\right) \,\,\, .
 \label{c1}
\eeq
There may be however another fixed point, given by the value 
$g_2 = - \pi /(4(1+2C_v)\alpha)$, where we have 
\beq
\left. {\cal{Z}}_{{\rm int}}\right\vert_{g_2} =
\int Dh Dg \exp\left\{(1+2 C_v) \left(\Gamma [g^T] +
\Gamma [h]\right)\right\} ,
\label{j}
\eeq
and the corresponding Virasoro central charge is given by 
\beq
c_2  = {N \over 2} - N(N-1) +  2 \left[{(1+2C_v)N(N-1) \over
2(1+C_v)}\right] 
 = {N \over 2} + N (N-1) \frac{C_v}{1 + C_v} \,\,\,. 
\label{c2}
\eeq
We have now to choose the parameter $\alpha$ and secondly to identify the
fixed points in the two ranges $N>2$ and $N<2$. 

Let us first see how the above points can be settled in the well-understood 
case $N > 2$. Since in this region the model is asymptotically free in the
ultraviolet and massive otherwise, no non-trivial fixed point is
expected. Therefore one should choose $g = \infty$ in (\ref{c}) since then
the integration over the gauge fields leads to constraints that eliminates
all the degrees of freedom yielding a $c = 0$ theory. 
This suggests the choice $\alpha=1$ ({\it i.e.}\ the gauge
invariant regularization) for which we have an infrared fixed point in
the strong coupling limit $g_1 \rightarrow \infty$. Plugging $C_v = N-2$
in the corresponding expression (\ref{c1}) of the central charge, we
find indeed the
correct value $c_1 = 0$. Notice that with this choice of $\alpha$, the
value of $g_2$ is negative and does not correspond to a physical fixed
point, because this is incompatible with the unitarity of the
GN model for $N > 2$.

The GN model at $N=2$ is somewhat special since it is equivalent to
the massless conformally invariant Thirring model with $c=1$ for 
$-{4 \over \pi} \le g < \infty$. At $N=2$ one has $C_v = 0$
and retaining $\alpha = 1$ as before, one finds that $g_1 = \infty$
still corresponds to a theory with $c_1 = 0$ which is consistent
with approaching $N=2$ from above out of a massive regime.
The other fixed point is now located at the end of the line where
$c=1$, {\it i.e.}\ $g_2 = -{4 \over \pi}$ and $c_2 = 1$ which
seems to be more appropriate for the approach to $N=2$ from
below out of the massless regime.

Let us now discuss the fixed points in the regime $N<2$ assuming 
that both formulas $\alpha = 1$ and $C_v = N - 2$ also apply here. The GN
model is generally non-unitary in this regime, a fact which opens the 
possibility to consider also a fixed point with a negative value of
the coupling constant. With the above choice of $\alpha$ and $C_v$, 
the central charge for the fixed point $g_1$ is identically
zero also for $N<2$. Let us consider the second critical value of the
coupling constant, {\it i.e.}\ $g_2$.  Note that $g_2$ is negative for 
$N >{3 \over 2}$ but positive otherwise and therefore also 
compatible with the statistical interpretation as the random-bond 
Ising model for $N \to 0$. The corresponding value of the ultraviolet
central charge is then given by (\ref{c2}), {\it i.e.}\ upon inserting
$C_v = N-2$
\beq
c_{{\rm uv}}  = \frac{N (2 N -3)}{2} \, .
\label{ccc}
\eeq
The above formula does not seem to apply to the case $N=1$:
in fact it predicts $c_{{\rm uv}}  = - 1/2$, but instead we
expect to find in this case $c_{{\rm uv}} = c_{{\rm ir}} = 1/2$ for the
simple reason that it is impossible to construct a quartic fermionic
interaction with only one Majorana fermion. The reason of this
mismatch seems somehow interesting: first notice that using (\ref{c2}) 
the variation of the central charge from the short to the large 
distances reads  
\beq
\Delta c \,=\,N (N-1) \frac{C_v}{1 + C_v} \,\,\,.
\label{variation}
\eeq
It is a general result of two-dimensional quantum field theories 
that such variations of central charges satisfy the sum rule
\cite{ctheorem} 
\beq
\Delta c = \frac{3}{4 \pi} \int d^2x \mid x\mid^2 \langle \Theta(x) 
\Theta(0) \rangle \,\, ,
\label{sumrule}
\eeq
where $\Theta(x)$ is the trace of the stress-energy tensor. This 
operator is proportional to the quartic interaction term in the 
Lagrangian (\ref{GNlag}) (the proportionality constant being the 
$\beta(g)$ function of the model). Therefore, the term $N (N-1)$ 
in (\ref{variation}) has a pure combinatorial origin. 
This factor itself of course vanishes for $N=1$. However, after
pulling out this combinatorial term, what is left in
(\ref{variation}) or in (\ref{sumrule}) may be interpreted as ``$\Delta c$
per unit of replica'', {\it i.e.}\ a quantity which still depends on 
$N$ but which has lost any reference to the color indices of the theory.
Let us denote it by $D(N)$. For $N=1$ this quantity can be easily computed
by means of the Green function of the free fermion resulting in a 
(logarithmically) divergent expression 
\beq
D(1) \sim \int d^2x \, \mid x\mid^2 \frac{1}{\mid x\mid^4} \,\,\, .
\eeq
For $N \neq 1$ we expect that the divergence is cured by the 
presence of interactions so that we expect $D(N)$ to be a finite
quantity\footnote{At $N=2$ it is the vanishing of the $\beta$ function 
which is responsable for the vanishing of $D(2)$.}. The above
considerations suggest then for $N=1$ there may be a sort of anomaly which
conspires to give a non-zero value for $\Delta c$. In this case the 
correct procedure might be to set $N=1$ in (\ref{c2}) before inserting 
the expression of $C_v$, {\it i.e.}\ to use a regularization expression 
for $D(N=1)$. 

\resection{Conclusions}
In summary, in this paper we have applied massless FFs to the critical regime
of the random-bond Ising model. Compared with the perturbative approach, the 
energy correlation function obtained with this method is well behaved in the 
whole range of scales. The technical difficulties encountered in the
computation of higher-particle FFs can possibly be circumvented by
using methods similar to the ones used in \cite{MeSm,Kar}, where FFs for
some non-diagonal $S$-matrices have been computed. One motivation of
our approach is to open the way for a more deep understanding of
the ultraviolet behavior of the massless flow of the Ising model
from its pure to its disordered regime and, more generally, for all
the $O(N)$ GN models with $N < 2$ which appear in the replica approach
of the original random problem. To this end we have also pointed out
a possible way of reaching the ultraviolet fixed point by a mapping of
the GN action to a WZW model, relating the strong and weak coupling 
regimes. This approach predicts the presence of a non-trivial 
fixed point with central charge given for $N \neq 1$ by (\ref{ccc}). 
The lacking of a sound mathematical definition for the relevant 
quantities of the group $O(N)$ for $N < 2$ makes it highly interesting 
to have independent information on this issue. In this respect, the most
natural approaches are those based on the Thermodynamical Bethe Ansatz
\cite{TBA} or the afore-mentioned $c$-theorem sum rule \cite{ctheorem}. 
The application of the two approaches seems however presently
obstructed by technical difficulties related both to the non-diagonal 
nature of the $S$-matrix and to the subtle problem of how to take the 
analytic continuation of the mathematical expressions for continuous values
of $N$ in the range $N < 2$ (this is particularly severe for a
Thermodynamical Bethe Ansatz approach). The solution of these problems
together with the massless FF approach proposed here may give
interesting non-perturbative information in the field of disordered 
systems.

\vskip 2 cm
\noindent
{\it Acknowldegments:} 
We would like to thank A.\ Ludwig, A.\ Lugo, E.\ Moreno and
M.S.\ Narasimhan for useful discussions.  D.C.C.\ is grateful to
the ICTP for hospitality and to CONICET and Fundaci\'on Antorchas,
Argentina for partial financial support. G.M.\ is partially supported
by the Istituto Nazionale di Fisica Nucleare.
              
\vfill
\end{document}